\documentclass[12pt]{iopart}
\usepackage[dvips]{graphicx}% Include figure files
\usepackage{dcolumn}% Align table columns on decimal point
\usepackage{bm}% bold math
\usepackage{subfigure}
\usepackage{hyperref}
\usepackage{array}

\begin{document}

\title[]{Influence network in Chinese stock market}

\author{Ya-Chun Gao$^{1}$, Yong Zeng$^{1}$, Shi-Min Cai$^{2}$}

\address{$^{1}$Department of Economics and Management,
     University of Electronic Science and Technology of China, Chengdu Sichuan, 611731, PR China}
\address{$^{2}$ Web Sciences Center, University of Electronic
Science and Technology of China, Chengdu Sichuan, 611731, PR China}

%\eads{\mailto{kuangben@mail.ustc.edu.cn} (Y.-C. Gao), \mailto{zengy@uestc.edu.cn} (Y. Zeng), \mailto{shimin.cai81@gmail.com} (S.-M.Cai)}

\eads{\mailto{shimin.cai81@gmail.com} (S.-M.Cai)}

\begin{abstract}
In a stock market, the price fluctuations are interactive, that is, one listed company can influence others.
In this paper, we seek to study the influence relationships among listed companies by constructing
a directed network on the basis of Chinese stock market. This influence network shows distinct
topological properties, particularly, a few large companies that can lead
the tendency of stock market are recognized. Furthermore, by analyzing the subnetworks of listed
companies distributed in several significant economic sectors, it is found that the influence relationships
are totally different from one economic sector to another, of which three types of connectivity as
well as hub-like listed companies are identified. In addition, the rankings of listed companies
obtained from the centrality metrics of influence network are compared with that according to
the assets, which gives inspiration to uncover and understand the importance of listed companies
in the stock market. These empirical results are meaningful in providing these topological properties
of Chinese stock market and economic sectors as well as revealing the interactively influence relationships
among listed companies.
\end{abstract}

%\vspace{2cm}
%{\it Keywords}: Hierarchical structure, financial price fluctuation, multifractal scaling, financial market.
\maketitle

\section{Introducation\label{1}}

In modern portfolio theory, risk diversification is the most essential issue,
which involves the understanding of clustering behavior and risk contagion
of the assets in a portfolio. Thus, in a stock market, the price fluctuate
of a listed company's asset (i.e., stock) is parallel to others or
interactively influenced by others. The widely used cross-correlation
analysis is an important measurement to investigate
the interactive relationships between pairs of stocks for understanding
the dynamic mechanics in complex economic system.
For example, the random matrix theory (RMT) suggests the eigenvalues and
corresponding eigenvectors of the cross-correlation matrix of price fluctuations
are relevant to clustering behavior and economic sector division
(or taxonomy) of stocks \cite{Plerou1999,Laloux1999,Jeong2005,Jiang2012}.
Meanwhile, with the development of complex network theory,
diverse cross-correlation based stock networks are proposed to
describe the interactive relationship, such as minimum
spanning tree (MST)~\cite{Mantegna1999,Bonanno2003,Onnela2003,Onnela2004},
planar maximally filtered graph (PMFG)~\cite{Tumminello2005},
and threshold networks (TN)~\cite{Kenett2010,Cai2010,Gao2013}, etc. Especially,
the clustering behavior of stocks can be
well associated with the communities scratched from these
stock networks via complex network measurement.

In order to evaluate risk contagion, a lot of works have been devoted to
analyze the influence relationships from directed network perspective.
Kenett \emph{et al.} \cite{Kenett2010} introduced
the measurement of partial correlation to construct TN and PMFG of listed companies and uncover the dominating ones in a stock market.
The Engle-Granger method~\cite{Granger1987} is an alterative way to obtain the asymmetric
influences (i.e, Granger causality) among listed companies. For example, Yang \emph{et al.}~\cite{Yang2014}
constructed directed cointegration network of global stock markets
based on Engle-Granger cointegration test, and presented ranking analysis of nodes to distinguish
their importance. Besides, time-dependent cross-correlation \cite{Kullman2002,Toth2006,Lee2011} is also applied
to determine the linking direction between a pair of listed companies due to
the time shift of maximum correlation. If the time shift is non-zero,
the``pulling" effect is assumed to exist in these listed companies.

As an important emerging market, the Chinese stock market possesses unique properties, such as stronger
cross-correlations and less market efficiency \cite{Gao2012}.
There are few works involved the unidirectional influence relationship \cite{Mai2014,Tu2014}.
However, these results are obtained based on daily stock returns, thus may be debatable under the consideration of efficient market.
In this paper, we mainly focus on the risk contagion in Chinese stock market, by constructing a directed influence network on
the basis of time series of minute-by-minute price fluctuations
with the time-dependent cross-correlation method, which is well behaved in American stock market\cite{Kullman2002}.
Unlike previous literatures, we analyzed not only the global topological structure,
but also the subnetworks of
a few significant economic sectors in aim to
explore the unique economic structure of China.
Empirical results reveal three types of connectivity
involving with the intra-sector's influence relationship.
We also compare several measurements of node's centrality
to find out available characterization of the importance of
listed companies in this influence network.
The findings provide intriguing information
about the topological properties of Chinese stock market
and give important hint about risk contagion in portfolio management.

\section{Materials and Methods \label{2}}

\subsection{Data sets}
%%%%%------------Table 1---------------%%%%%%%%%%%%
\begin{table}
\center
\caption{\label{tab:table1}
Number of stocks from each economic sector in the data set.}
\begin{tabular}{lclc}
\hline \hline
Sector & Number & Sector & Number\\
\hline
Finance & 22 & Construction & 23 \\
Mining Industry & 30 & Energy & 42 \\
Manufacturing & 418 & Real Estate & 58\\
Wholesale\&Retail & 59 & Transportation & 47 \\
Lodging\&Catering Service & 3 & Agriculture & 14  \\
Information Technology & 20 & Other Service & 1 \\
Lease\&Business Service & 8 & Utility & 2 \\
Science\&Technology Service & 1 & Healthcare & 1 \\
Public Management & 25 & Entertainment & 5 \\
\hline \hline
\end{tabular}
\end{table}

The data set consists of $N = 779$ stocks (i.e, listed companies) trading in Shanghai security exchange (SSE).
These stocks belong to 18 economic sectors, of which the name and size are shown in Tab.~\ref{tab:table1}.
The price fluctuations are sampled with minute frequency, which can \emph{quickly} respond to interactive
influence relationships among stocks. The duration is whole fiscal year of 2010,
totally including 242 trading days with 4 hours working time.
For the price fluctuation of each stock,
its return at time scale $\Delta t$ is obtained by
\begin{equation}
r_{\Delta t}(t)=\frac{\ln [p(t)]}{\ln [p(t-\Delta t)]}.
\end{equation}
We set $\Delta t$=1 minute because larger $\Delta t$ may smear out of the maximum.
and $r_{\Delta t}(t)$ is denoted by $r(t)$ for simplicity.

\subsection{Time-dependent Cross-Correlation}

To evaluate the interactive influence relationships among stocks,
their time-dependent cross-correlations are calculated.
Within a trading day $T$, the correlation between stocks
$i$ and $j$ can be calculated as
\begin{equation}
C_{i,j}^{T}(\tau)=\frac{\langle r_{i}(t)r_{j}(t+\tau) \rangle-\langle r_{i}(t) \rangle \langle r_{j}(t+\tau) \rangle}{\sigma_i\sigma_j},
\end{equation}
where $\sigma_i$ and $\sigma_j$ are the standard deviation of $r_i$ and $r_j$
and the parameter $\tau$ in $C_{i,j}^{T}(\tau)$ is time shift.
Changing $T$, the $C_{i,j}^{T}(\tau)$ are then
averaged over trading days to filter the dairy effect~\cite{Harris1986,Admati1988,Ekman1992},
and the mean value is denoted by $C_{i,j}(\tau)$.

%%%%%%%%%%%%----Figure 1-----------%%%%%%%%%%%%%%%%%%

\begin{figure}
\center
\includegraphics[width=9cm]{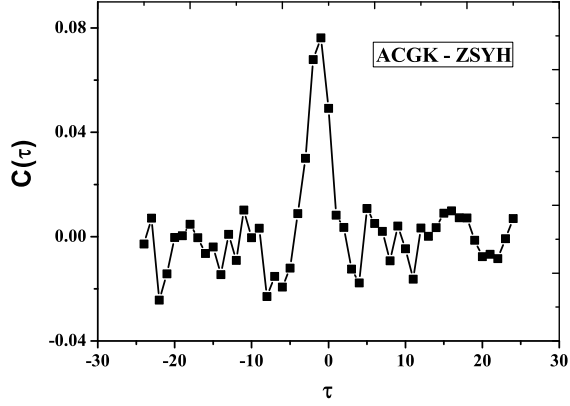}
\caption{\label{fig:1}(Color online) The time dependent cross-correlation between stock ACGK and ZSYH as a function of
time shift $\tau$. The maximum value appears at $\tau_{max}(i,j)=-1$, indicating that stocks ZSYH influences
stock ACGK in their price fluctuations. All the abbreviations for companies are listed in Tab.~\ref{tab:table4}.}
\end{figure}

With a various value of $\tau\in [-100,100]$, the corresponding $C_{i,j}(\tau)$
are then obtained, of which the maximal value is selected, denoted as $C_{max}(i,j)$,
and its related time shift as $\tau_{max}(i,j)$. For example, as shown in Fig.~\ref{fig:1},
$C_{i,j}(\tau)$ between stocks $i$ (ACGK) and $j$ (ZSYH) changes with various $\tau$,
where $C_{max}(i,j)=0.07$ is obtained at $\tau_{max}(i,j)=-1$. It suggests that stock $j$ influences
stock $i$ in their price fluctuations. Besides, to differentiate from $C_{max}(i,j)$ to noise, the
parameter $R(i,j)$ is measured as the ratio of $C_{max}(i,j)$ and the noise strength
defined as the variance of all correlation values with time shift from the peak larger
than 10 min because the largest peak width is 6 min.

\subsection{Influence network construction}

\begin{figure*}
\center
\subfigure[]{\includegraphics[width=6cm]{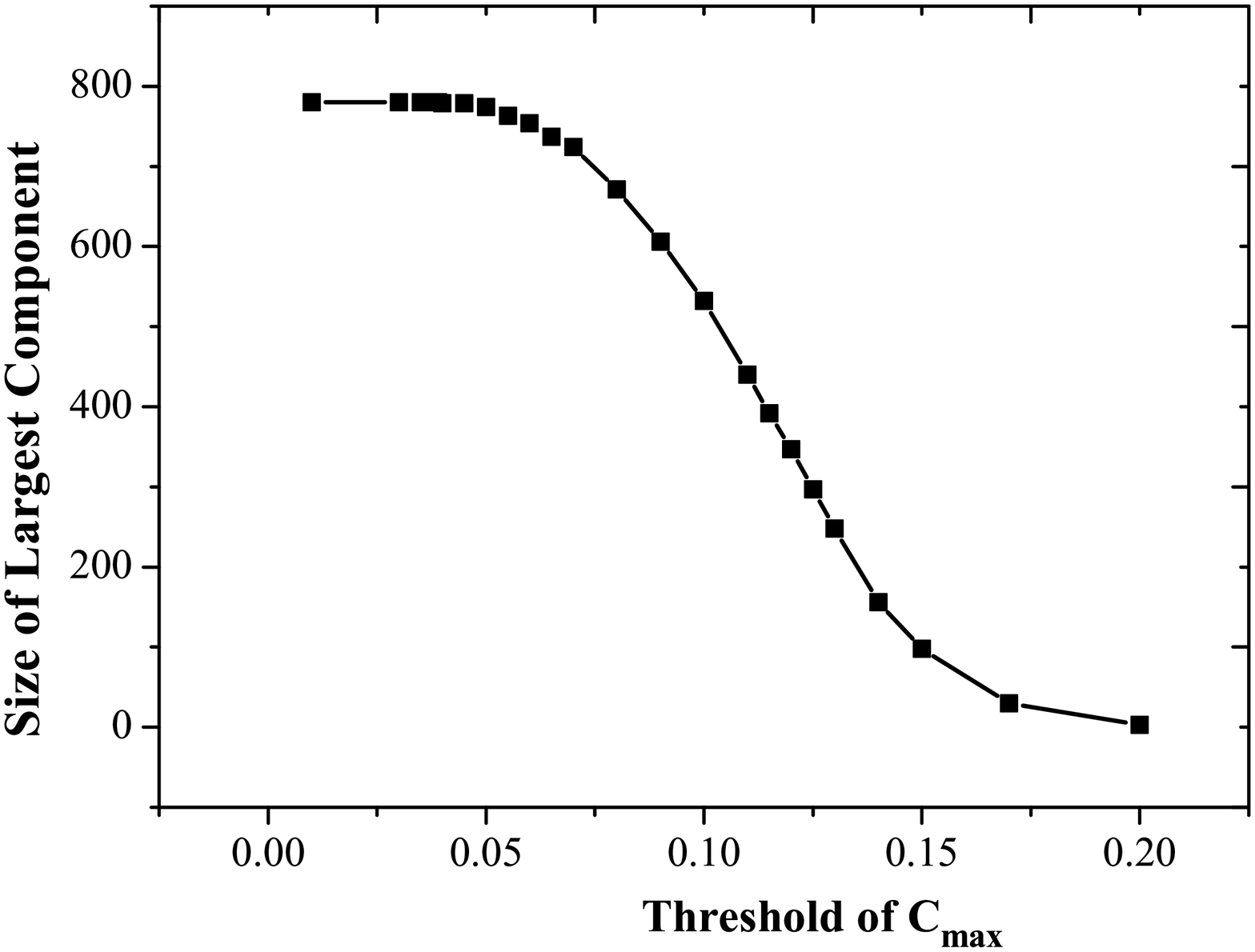}\label{fig:2a}}
\subfigure[]{\includegraphics[width=6cm]{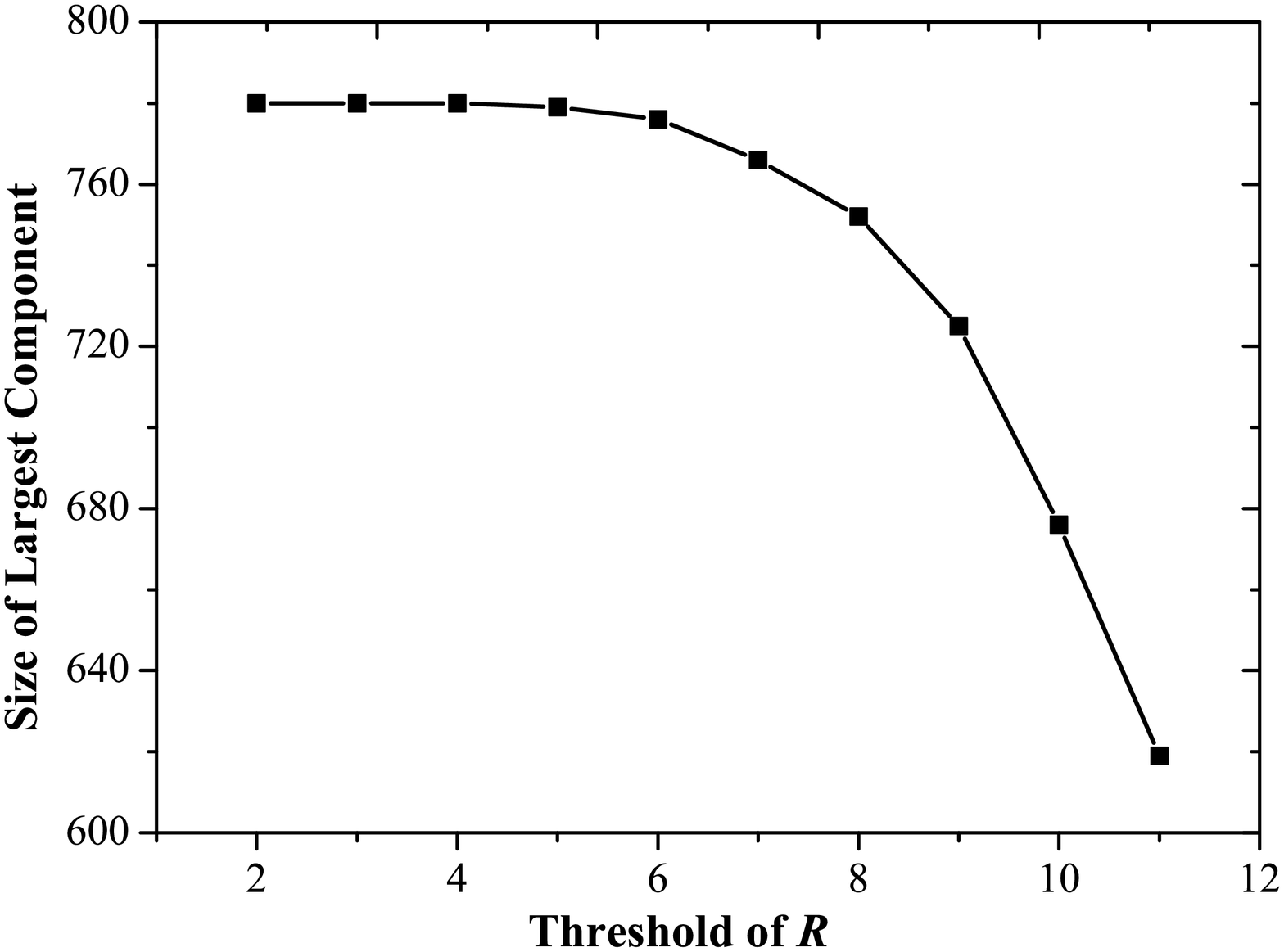}\label{fig:2b}}
\caption{The size of largest component of influence network versus various threshold of (a) $C_{max}$ \textbf{with $R=4$}, (b) $R$ \textbf{with $C_{max}=0.039$} .}
\label{fig:2}
\end{figure*}

With time-dependent cross-correlations, the influence relationships of all pairs of stocks can be quantitatively measured.
To construct a directed network describing influence relationships, we adopt the method proposed in \cite{Kullman2002},
which emphasizes that three parameters  $C_{max}(i,j)$, $|\tau_{max}(i,j)|$, and $R(i,j)$ should exceeded certain
threshold values simultaneously if the directed connection between stocks $i$ and $j$ exists.
It is obvious that the topological structure of influence network has a direct relevance to these thresholds.
Figure~\ref{fig:2} shows the size of largest component as a function of $C_{max}$ and $R$, respectively.
One can see that the size of largest component decreases whenever improving the threshold value of $C_{max}$ or $R$, because more links are filtered. And in both cases, there is a critical point when the full-connected network decomposes and the size of largest component decreases rapidly.
Based on percolation-based method~\cite{Cai2010,Gao2013},
the value of phase transition point from full connection to isolated components is $C_{max}\geq 0.04$ and
$R\geq 4$. Moreover, $|\tau_{max}(i,j)| \geq 1$ is required.

%%%%%%%%%%%%%%%%%%-----Figure 2-----------------%%%%%%%%%%%%%%%%%%%%%

In the influence network, link $L_{i,j}$ between stocks $i$ and $j$
are unidirectional, determined by the sign of $\tau_{max}(i,j)$.
If $\tau_{max}(i,j)<0$, the current price of stock $i$ is affected
by the previous one of stock $j$, denoting the link
direction from $i$ to $j$. Otherwise, the link is directed
to $i$ from $j$ if $\tau_{max}(i,j)>0$.
It should be pointed out that, in our network,
a directed link is set from $i$ to $j$ if $j$ influences $i$,
however, the reverse is also a feasible choice.
When $\tau_{max}(i,j)=0$ (i.e., the equal time cross-correlation), we recognize
the mutual influence as an external effect.
The price fluctuations of two stocks may be induced by the
trend of stock market or environmental variation
in the economic sector. Therefore, in this case, stock $i$ and $j$
aren't connected.

In order to further get rid of the noises that the maximum of the correlation
is attributed to occasional large values rather than a real association,
the fiscal year are divided into three periods, in each period
the $C_{i,j}(\tau)$ is calculated, according to which an adjacent matrix
is established, and only those links existed in all three periods
are considered in the network to ensure the robustness of the result.

\section{Empirical results}\label{3}
\subsection{Analysis of influence network}
\begin{figure}
\center
\includegraphics[width=9cm]{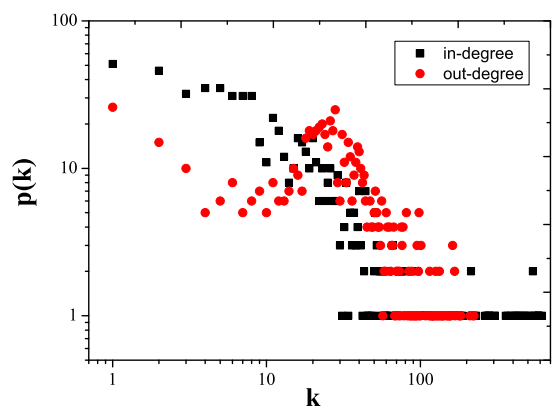}
\caption{\label{fig:3}(Color online) Distributions of in-degrees (black squares) and out-degrees
(red solid circles) of the influence network. The fat-tail both in the in-degree and out-degree distributions
suggests that there are hub-like nodes that strongly influence others.}
\end{figure}

\begin{figure}
\center
\includegraphics[width=9cm]{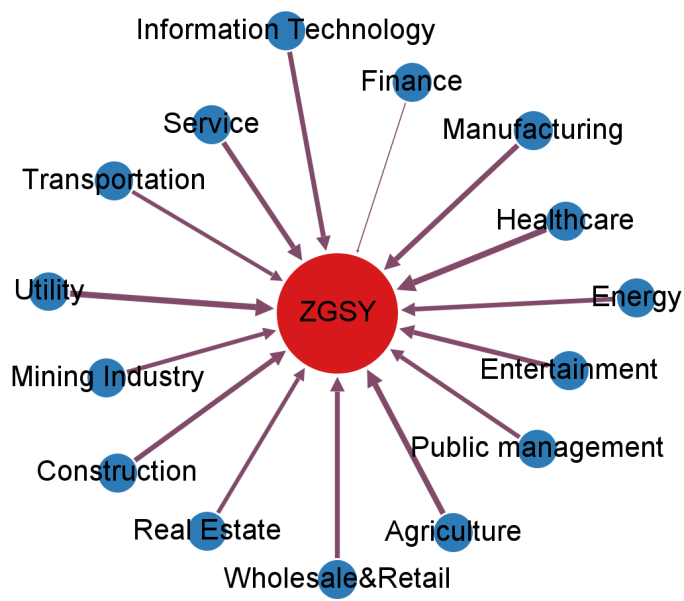}
\caption{\label{fig:4} (Color online) A visualization of ZGSY's influence
to 616 stocks distributed into 17 economic sectors.
The thickness of edges denotes the number of stocks
linked to ZGSY in a certain economic sector. Concretely,
The proportions in 18 economic sectors are 13/14 in Agriculture,
21/30 in Mining Industry, 359/418 in Manufacturing, 32/42 in Energy,
19/23 in Construction, 47/59 in Wholesale$\&$Retail,
30/47 in Transportation, 11/13 in Service (including 4 types),
16/20 for Information Technology, 39/58 in Real Estate, 4/5 in Entertainment.
in 2/2 Utility, in 1/1 Heathcare, 18/25 Public Management}
\end{figure}

The resulting influence network has dense edges, with the average degree high to 34.84.
Figure~\ref{fig:3} shows the distributions of in-degree and out-degree, respectively.
Both of them approximately decays in an exponential way when the degree $k$ is at a small scale.
Nevertheless, the fat tail both in the in-degree and out-degree distributions
reveals there are some hub-like nodes in the influence network. In other words,
a few huge stocks can strongly influence, or even control the trend of Chinese stock
market. In Tab.~\ref{tab:table2}, it shows that the top-10 stocks with the highest in-degree are mainly distributed
in Mining Industry and Finance sectors, and almost affect the whole stock market.
For example, ZGSY, the largest listed company in China, influences the
more than 600 stocks in all economic sectors, as shown in Fig.~\ref{fig:4}. However, it is
interesting that the majority of all the economic sectors are influenced, except for that
of finance, only a fraction of 3/22 are linked to ZGSY.

%%%%%%%%%%%%%%%%%%%%%%%%%%%---Table 2--------------%%%%%%%%%%%%%%%%%%%

\begin{table}
\center
\caption{\label{tab:table2}
Top-10 companies with the highest in-degrees.}
\begin{tabular}{llcllc}
\hline \hline
 Sector&Company&In-degree&Sector&Company&In-degree \\
\hline
Mining Industry & ZGSY& 616&Finance & ZGTB & 594 \\

Finance & ZSYH & 585&Finance & JTYH & 580 \\

Finance & HXYH & 578&Finance & ZGRS & 551 \\

Finance & PFYH & 542&Mining Industry & SHE & 538  \\

Finance & XYYH & 538&Finance & BJYH & 511  \\
\hline \hline
\end{tabular}
\end{table}

%%%%%%%%%%%%%%%%-----Figure 3---------------%%%%%%%%%%%%%%

%%%%%%%%%%%%%%%%-----Figure 4---------------%%%%%%%%%%%%%%

\begin{figure}
\center
\includegraphics[width=9cm]{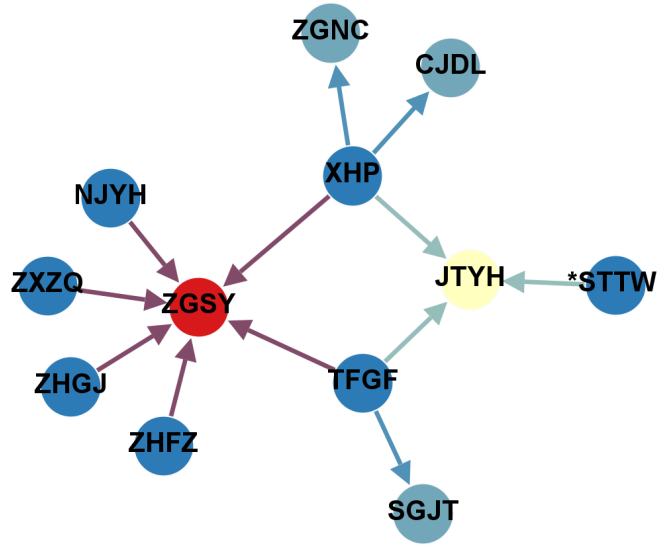}
\caption{\label{fig:5} (Color online) The connectivity among the top 50 companies with the highest in-degrees.
Few edges existed in these core-like influence network.}
\end{figure}

In addition, we also pay attention to the interactive
influence relationships among stocks with top-50 in-degree.
Figure~\ref{fig:5} shows that there are only 12 directed
connections, which suggests that these stocks are relatively
independent, that is, their price fluctuations are
parallel to each other. Nevertheless, ZGSY
still plays an important role in this core as its
in-degree is 6, equalling half of total connections.

%%%%%%%%%%%%%%%%-----Figure 5---------------%%%%%%%%%%%%%%

%%%%%%%--------------------9-18-----------%%%%%%%%%%%%%%%%%%%%%%%%%

\begin{figure}
\center
\includegraphics[width=9cm]{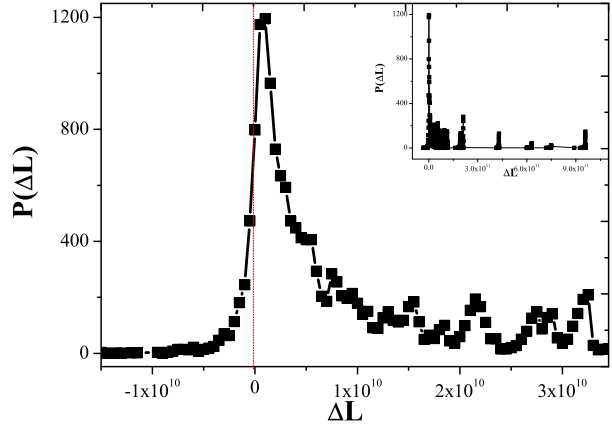}
\caption{\label{fig:6} (Color online) Distribution of $\Delta L_{ij}$ for all links.
Inset is the full view of the distribution while the large image zooms in
around the peak. The red dots mark the position of $\Delta L_{ij}=0$.}
\end{figure}

Furthermore, we have noticed that those nodes with high in-degrees (namely more influence)
have high capitalization. The positive correlation between influence and capitalization
has been studied by Lo and MacKinlay \cite{MacKinlay1990} with weekly return data.
To observe this effect in high-frequency return data, we calculate the difference of
the assets of the two connected nodes $i \rightarrow j$ as \cite{ZhouWX2014}
\begin{equation}
\Delta L_{ij}=L_j-L_i
\end{equation}
where $L$ represents equity capital, obtained by averaging equity capitals of the beginning and the end of 2010. Figure~\ref{fig:6} shows the distribution of all $\Delta L$ values for the whole network. It can be found that the peak locates
at $\Delta L>0$ rather than zero, and the shape of the peak is asymmetric, as the right side is more fatter.
These properties clarify that smaller capitalized listed companies tend to be affected by bigger capitalized listed companies but not vice versa, which is in accordance with previous study,
thus confirming the validity of this influence network.
Take a note of the inset, the right tail tend to
be growing, which is related to the fat tail of in-degree distribution.

In the above discussion, the most influential stocks are concerned.
Beyond that, we also analyze the most influenced stocks represented by higher
out-degree to better understand influence network. As shown in Tab.~\ref{tab:table3},
the top-10 stocks with the highest out-degree are displayed, along with their economic sectors.
One can see that they are completely different from those most influential ones.
Compared to Tab.~\ref{tab:table2}, their values of out-degree are much lower than that of
in-degree, which suggests that these stocks are influenced by only a portion of other stocks,
and these most influenced stocks are distributed to more diverse economic sectors, such as
Manufacturing (5), Real Estate (2), Energy (1), Transportation (1), and Wholesale$\&$Retail (1).
Nevertheless, it is easy to understand the difference because in stock market
these influential stocks are able to pull others via cascading effect of network
but those influenced stocks aren't ensured to be attracted by all other ones.

%%%%%---Table 3--------------%%%%%%%%%%%%%%%%%%%
\textbf{
\begin{table}
\center
\caption{\label{tab:table3}
Top-10 companies with the highest out-degrees.}
\begin{tabular}{llcllc}
\hline \hline
 Sector&Company&Out-degree&Sector&Company&Out-degree \\
\hline
Manufacturing	&	SBGX	&	223	&	Real Estate	&	ZFGF	&	220	\\
Manufacturing	&	BXGF	&	210	&	Manufacturing	&	SHSC	&	183	\\
Energy	&	GDDL	&	182	&	Transportation	&	TJHY	&	178	\\
Manufacturing	&	MYL	&	168	&   Wholesale\&Retail	&	BHC	&	167	\\
Manufacturing	&	FRYY	&	167	&	Real Estate	&	SQF	&	162	\\
\hline \hline
\end{tabular}
\end{table}}

\subsection{Analysis of subnetworks in economic sectors}

%%%%%%%%%%%%%%%%%%-------Figure 7-----------------%%%%%%%%%%%%%%%%%%%%%%%%%%
\begin{figure}
\center
\subfigure[]{\includegraphics[width=6cm]{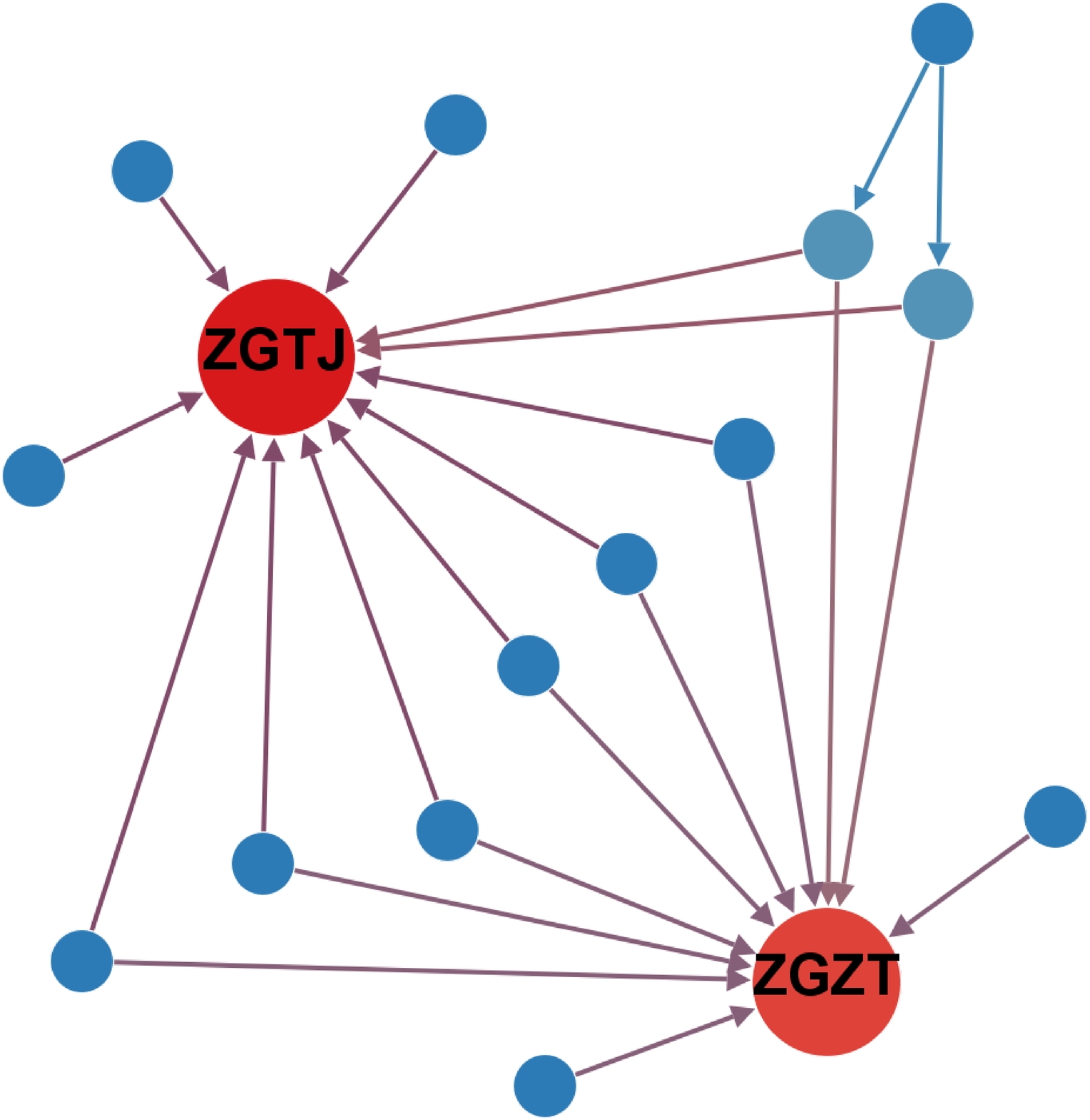}\label{fig:7a}}
\subfigure[]{\includegraphics[width=6cm]{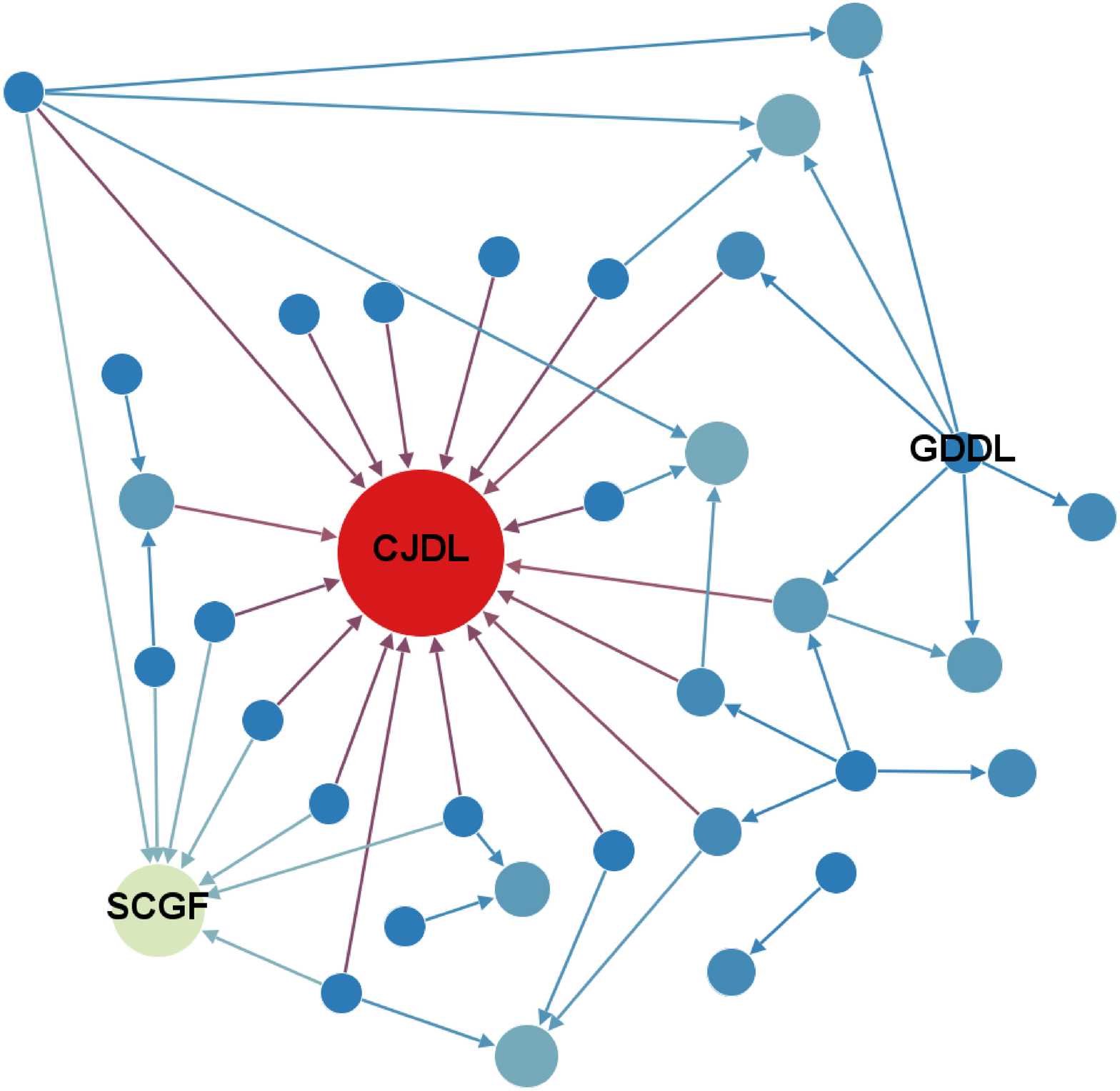}\label{fig:7b}}
\subfigure[]{\includegraphics[width=6cm]{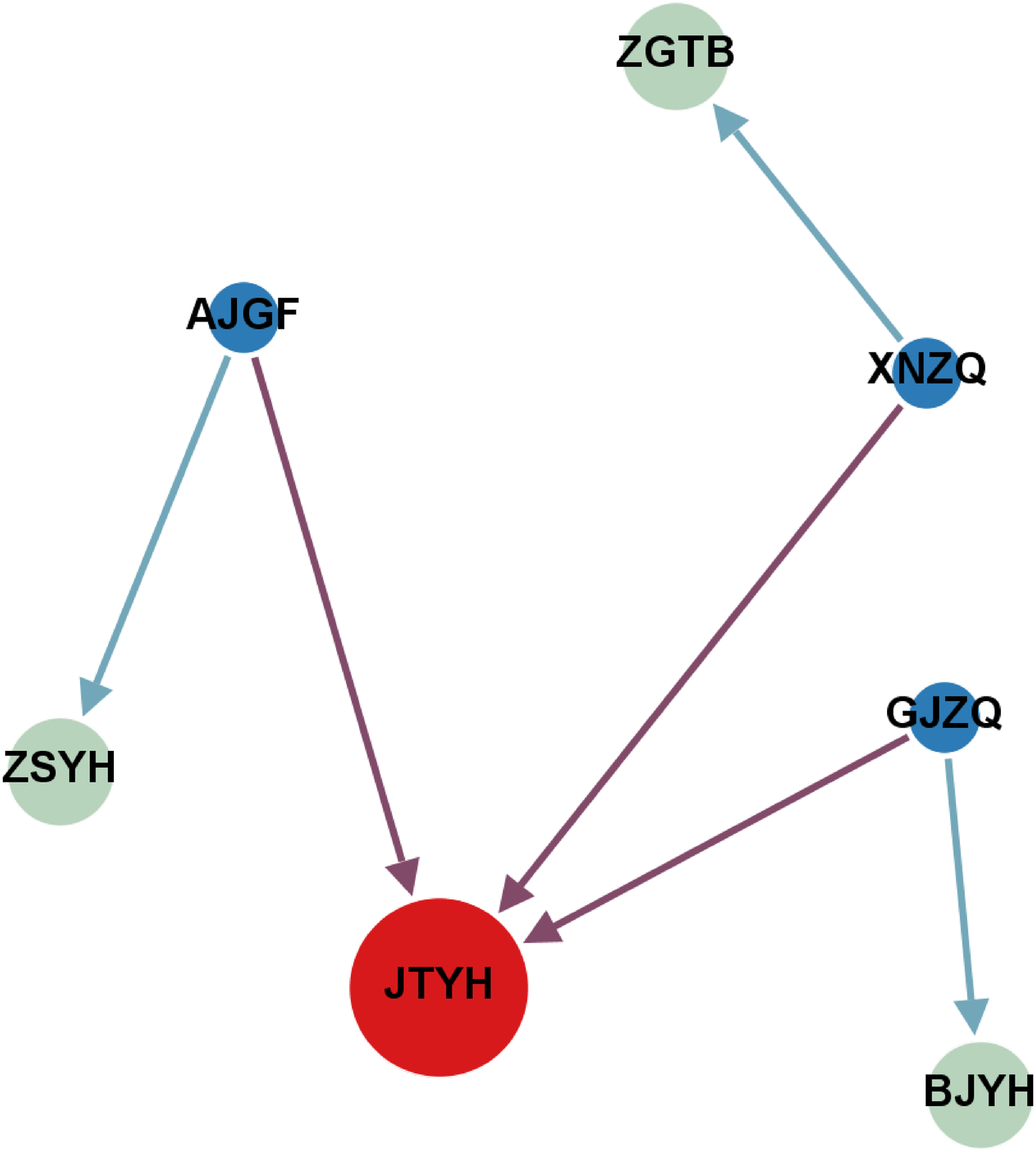}\label{fig:7c}}
\subfigure[]{\includegraphics[width=6cm]{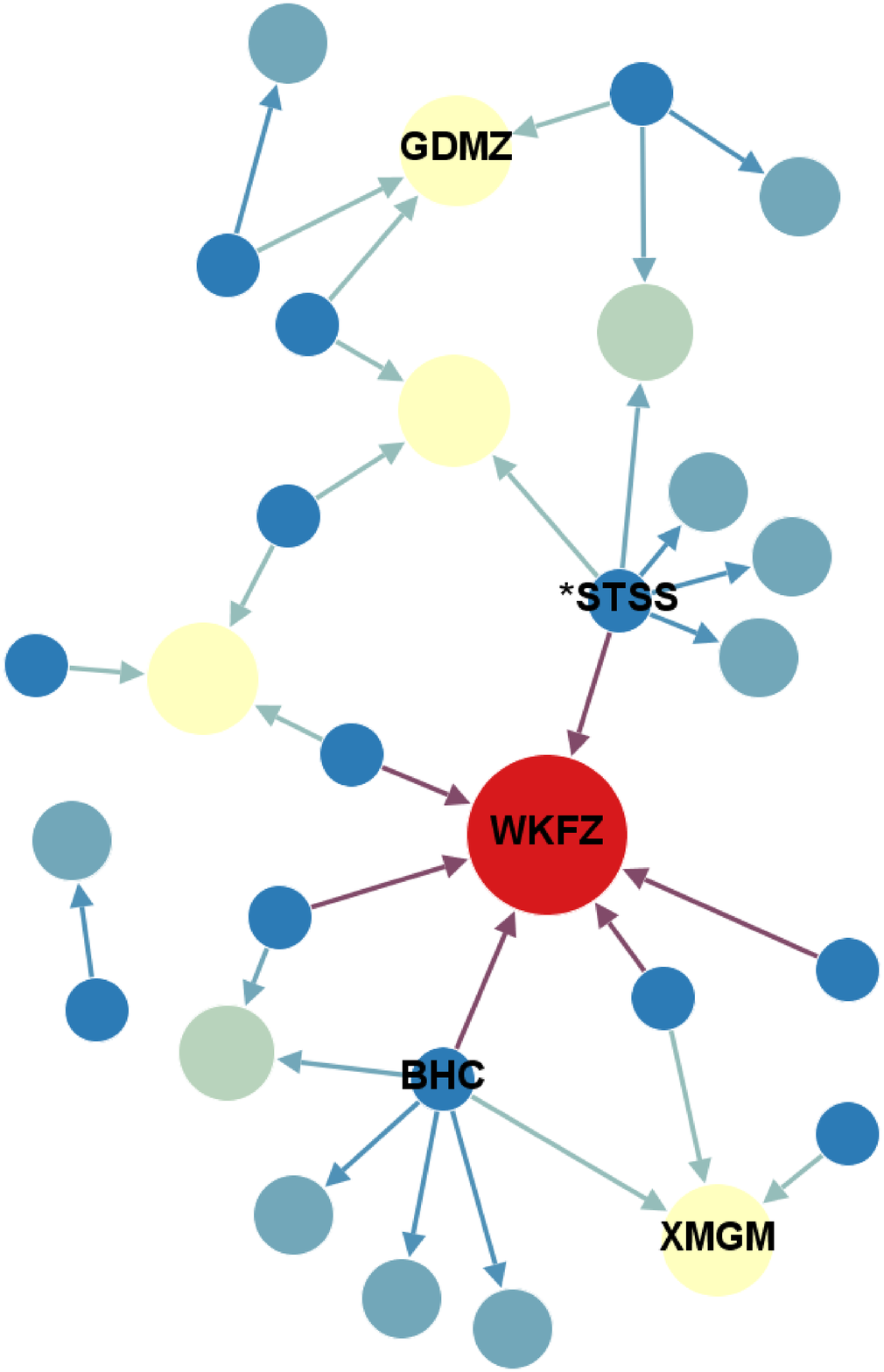}\label{fig:7d}}
\subfigure[]{\includegraphics[width=6cm]{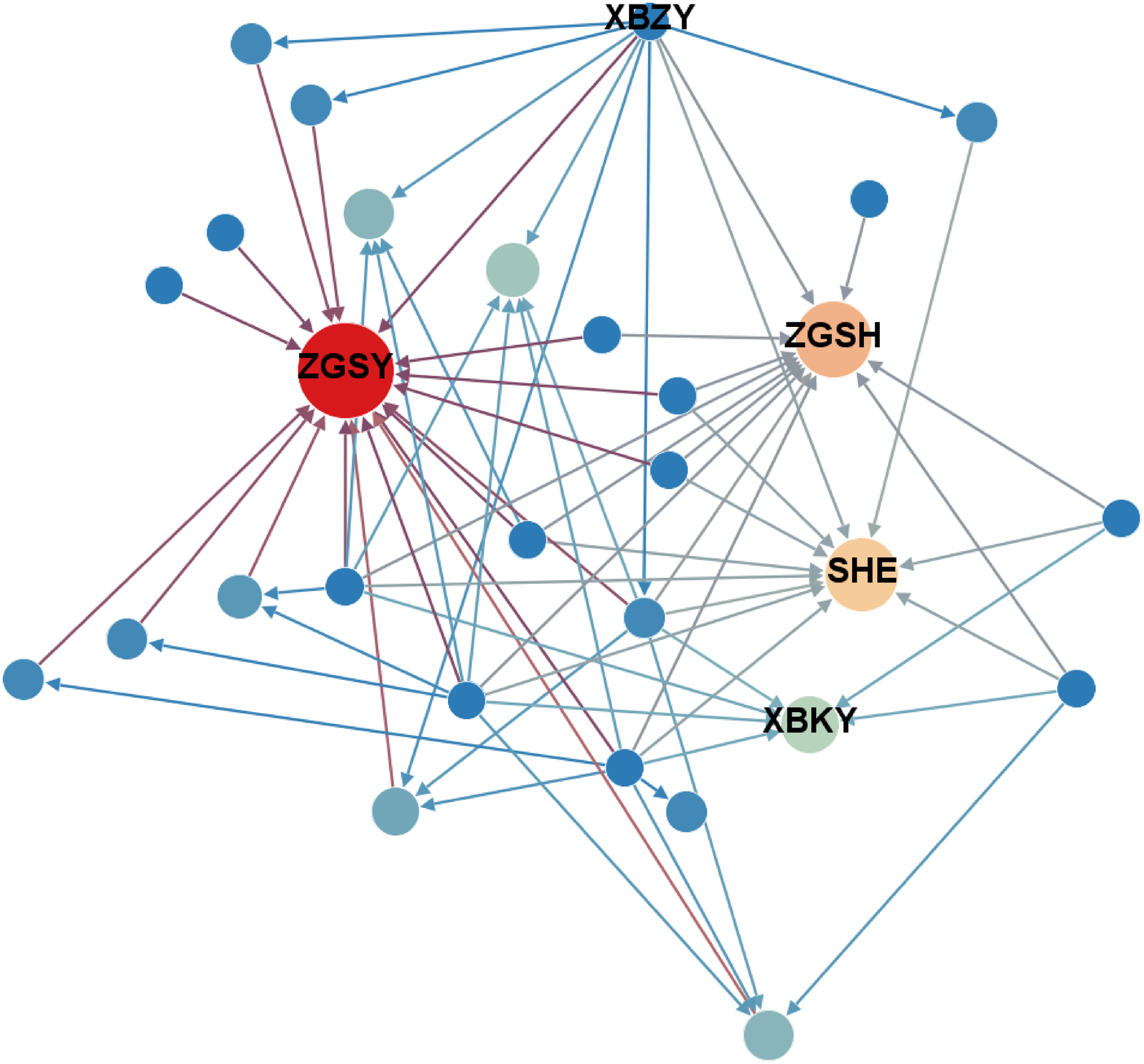}\label{fig:7e}}
\subfigure[]{\includegraphics[width=6cm]{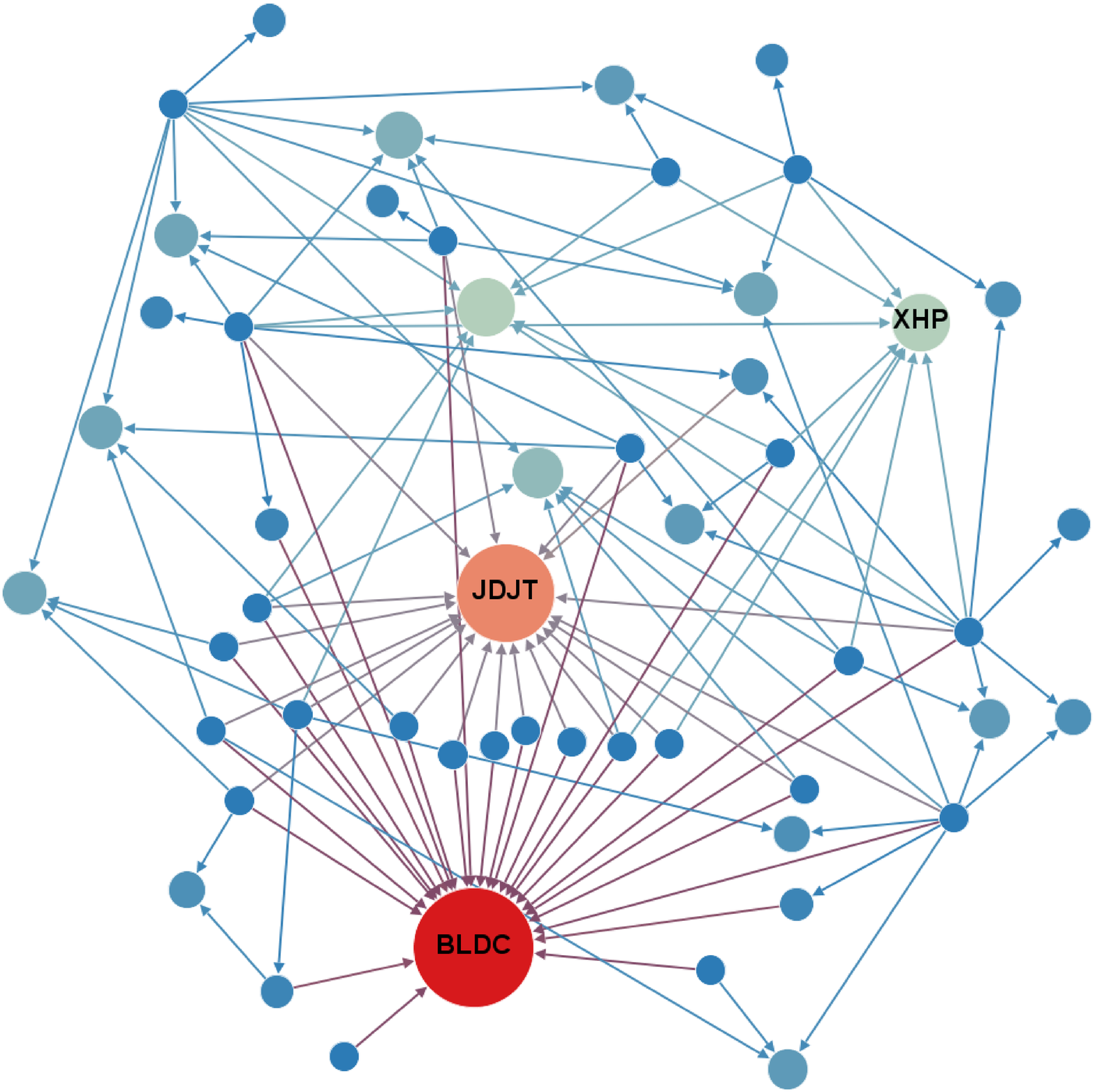}\label{fig:7f}}
\caption{(Color online) The subnetwork for 6 economic sectors of (a)
Construction, (b) Energy, (c) Finance, (d) Wholesale$\&$ Retail,
(e) Mining Industry, (f) Real Estate. The color and size of the solid circles
correspond to the in-degrees. Three configuration are recognized:
little connection with hub nodes (e.g. subfigure (a) and (b)); little
connection and no hub nodes (e.g. subfigures (c) and (d));
much more intra-connection with hub nodes (e.g., subfigures (e) and (f)).
Note that the most vulnerable nodes are also marked in
addition to the most essential ones.}
\label{fig:7}
\end{figure}

We have given an overview investigation of influence network at whole scale,
however, information of interactive influence relationships
in intra-sectors are required to be probed. Based on taxonomy of Chinese stock market,
we obtain a series of subnetworks from whole influence network. Figure~\ref{fig:7}
shows 6 significant economic sectors, such as Construction, Wholesale$\&$Retail,
Finance, Mining Industry, Energy, Real Estate. It can be found that
the connection configurations are different from each other, on
the basis of which the sectors can be classified into three types:

(1) There are few edges inside the sector, but hub nodes are apparent.
For instance, in the sector of Construction (Fig.~\ref{fig:7a}), there
are two key nodes of highest in-degrees, corresponding to the
industry heavyweights, ZGTJ and ZGZT.
The same properties can also be found in the sector of Energy.
As shown in Fig.~\ref{fig:7b}, CJDL is the largest listed company in the sector of Energy,
as well as the second largest one SCGF.
The most susceptible vertex is GDDL.

(2) The stocks rarely interact with others in the same sector,
and also there are no apparent hub-like nodes as the distribution
of in-degrees is approximately homogenous, such as Finance (Fig.~\ref{fig:7c}) and Wholesale $\&$
Retail (Fig.~\ref{fig:7d}). Concretely,
in Fig.~\ref{fig:7d}, the relatively important nodes, denoted by
red circle, is WKFZ (providing metals and metallurgical
raw materials), two other susceptible nodes, *STSS and BHC,
are also tagged. Although in the global network analyzed above,
8 of the top-10 in-degree nodes are financial stocks,
which can affect a large quantity of nodes in the whole network,
they barely influence each other, however. Furthermore,
it is interesting that the Finance sector is insensitive
to other sectors, yet they impact all the other sectors.

%%%%%%%----Figure 8-------%%%%%%%%%%%%%%
%\begin{figure}
%\includegraphics[width=3.2in]{fig8}
%\caption{\label{fig:8} (Color online) The influence relationships between financial sector and other %sectors.}
%\end{figure}

(3) The intra-sector influence relationships are much more considerable
compared with the first two classes, and the industry giants
can be observed easily from these subnetworks. In Fig.~\ref{fig:7e},
the Mining industry has four huge listed companies, ZGSY,
ZGSH, SHE, and XBKY.  Besides, the two observed
giants are BLDC and JDJT in the sector of Real
Estate (Fig.~\ref{fig:7f}).

\subsection{Analysis of node centrality}

%%%%%%%%%%%%%%---------------------Table 4-----------%%%%%%%%%%%%%%%%%%%%%%%%%%%%%%%
\begin{table}
\center
\caption{\label{tab:table4}
Similarity between enterprise value (represents by Equity capital, Total assets and ROA) and node centrality measurements in terms of Kendall's Tau coefficient.}
\begin{tabular}{ccccccc}
\hline \hline
	&	In-degree	&	PageRank	&	Eigenvector	&	Authority	&	Hub	&	Betweenness	\\
\hline
Equity	&	0.4072	&	0.4072	&	0.4108	&	0.4072	&	0.0767	&	-0.1905	\\
Assets	&	0.3772	&	0.3773	&	0.3805	&	0.3773	&	0.0594	&	-0.1938	\\
\hline \hline
\end{tabular}
\end{table}

It is a critical problem to evaluate the node importance in a directed network, and dozens of centrality metrics have been proposed, such as Betweenness Centrality (BC)~\cite{Freeman1977,Newman2005},
Eigenvector Centrality (EC)~\cite{Newman2010}, PageRank (PR)~\cite{Berkhin2005,Bryan2006},
Hub and Authority~\cite{Kleinberg1999}, which derive from diversely local topological
properties of influence network.
However, it remains an unsolved issue which is the appropriate centrality measurement which can reflect the economic importance in a financial network.
In this section, the ranking analysis of nodes based on these measurements are performed.
On the other hand, the nodes are also ranked due to the capitalization of listed company,
for which both total assets (including equity capital and liabilities) and
equity capital are considered, denoted by assets and equity, respectively.
The similarities of node ranking between assets and other centrality measurements
are then calculated in terms of Kendall's Tau (KT) coefficient~\cite{Kendall1938},
as well as equity. To keep our description in self-contained, we briefly introduce KT correlation.
For two sequences $\{x_i\}$ and $\{y_i\}$, $i=1,2,...,N$, the KT coefficient is given by
\begin{equation}
\tau=\frac{2}{N(N-1)}\Sigma_{i<j}sgn[(x_i-x_j)(y_i-y_j)]
\end{equation}
Here $sgn(x)=1$ for $x>0$, while $sgn(x)=-1$ for $x<0$, otherwise $sgn(x)=0$.
The result is shown in Tab.~\ref{tab:table4}.

One can see from Tab.~\ref{tab:table4} that all the KT coefficients of
equity are higher than those of assets. This is consistent with empirical
observations that the influence of a listed company is positively
related to its market capitalization in the equity market. More concretely,
we discuss the correlation between each centrality measurement and equity (or asset) as follows.

First, it is not surprising that BC is in negative correlation with assets.
BC of a vertex is defined as the frequency that it is in the shortest path between any two other vertices.
In the directed influence network, the shortest path between a pair of nodes
is asymmetric.
And the topological properties of the influence network have suggested that the important nodes are of large in-degrees and few out-degrees, therefore, their BC values are very small, even zero for those with zero out-degrees.

Second, Hub and Authority are two parameters of HITS. For a node,
its hub is determined by the authority of out-degree
neighbors, while its authority is confirmed by the hub of in-degree neighbors.
Thus, in the influence network, the authority of a node with larger asset is higher due to
its great number of in-degree neighbors, while its hub isn't greatly larger than those of other nodes
with less assets due to its smaller out-degree neighbors.
On the other hand, most nodes with less assets connect collectively to those with larger assets,
so no significant difference exists between their hub values.
These explains the poor performance of hub, and better performance of authority.

Third, both PR and EC can well indicate node's importance, suggested
by the higher KT coefficient shown in Tab.~\ref{tab:table4}. It
is comprehensible because the two measurements have similar idea that the
importance of a node depends not only on the numbers but also importance of its
in-degree neighbors. Although there are some arguments of the eigenvector of
a directed network~\cite{Newman2010}, it is practical for nodes with
high in-degree, which is suitable for the influence network in this paper.
Also of note is that the low value of the KT coefficient is in concerned
with the degree distribution. High in-degree nodes is of low out-degree,
and the less important nodes is uniformly out-degree distributed.

\section{Conclusion}
In this paper, in order to investigate the interactively clustering behavior of listed companies induced
by asymmetric market information, we have studied the influence network constructed from the
time-dependent cross-correlation of stocks' price fluctuations in Chinese stock market. The empirical
results can be concluded in three aspects.

From the distribution of the asset difference of all pairs of connected nodes,
the good performance of the network is verified
in revealing the influence relationships among listed companies.
However, the network is found to display singular topological properties
in the in-degree distribution, which can be attributed to the existence of
hub-like listed companies that can influence the majority of the Chinese stock market. The out-degree distribution, on the other hand, is more diverse.

In addition, the intra-sector influence relationship is also analyzed from subnetworks of
a few economic sector involved with Chinese economy. The topological
structure of the subnetworks differs among sectors in connectivity
and hub nodes. Three configurations are identified: Few edges with apparent
hub nodes as Construction; Few edges without hub nodes, such as Wholesale$\&$Retail
trades and Finance; Lots of links with apparent hub vertices, like Mining Industry,
Energy, and Real Estate. These results gives important information in price
fluctuations in the stock market, that is, they implies that the asymmetric
market information transferring from one economic sector to whole stock market
behaves diverse dynamic patterns. These may have significant applications
for portfolio management and risk diversification.

In order to figure out which algorithms can characterize critical nodes
in the influence network, we calculated the similarities between several
centrality measurements and assets of listed companies, which is
regarded as an indicator of their importance in Chinese stock market.
We found that the in-degree, PR, EC, as well as authority better characterize
the importance of listed companies, while BC and hub fail to.

\section*{Acknowledgments}
The authors acknowledge the support of National Natural Science Foundation of China (Grant No. 71472025) and
China Postdoctoral Science Foundation (Grant No. 2014M552350).

%\clearpage
 \begin{appendix}
\section*{APPENDIX}

\setcounter{table}{0}
\renewcommand{\thetable}{A\arabic{table}}
%\begin{appendix}
\nopagebreak[4]
%%%%%------------Table A1---------------%%%%%%%%%%%%
\begin{table}
\center
\caption{\label{tab:table5}
Company names and symbols mentioned in the article.}
%\begin{ruledtabular}
\begin{tabular}{ccl}
\hline \hline
Label \footnote{Label is the trading ticker of each security on sale in SSE.}	&	Symbol	&	Name\\
\hline
600234	&	*STSS\footnote{Companies with *ST (Special treatment) in the stock ticker is in abnormal financial situation for three consecutive years.}	
&	Guanghe landscape Culture Communication	\\
600550	&	*STTW	&	Baoding Tianwei Baobian Electric Co.	\\
600207	&	ACGK	&	Henan Ancai Hi-tech Co.	\\
600643	&	AJGF	&	Shanghai Aj Corporation	\\
600721	&	BHC	&	Xinjiang Baihuacun Co.	\\
601169	&	BJYH	&	Bank Of Beijing Co.	\\
600048	&	BLDC	&	Poly Real Estate Group Co.	\\
600083 & BXGF &Guangdong Boxin Investing \& Holdings Co. \\
600900	&	CJDL	&	China Yangtze Power Co.	\\
600781  & FRYY & Furen Pharmaceutical Group Co.\\
600310	&	GDDL	&	Guangxi Guidong Electric Power Co.	\\
600382	&	GDMZ 	&	Guangdong Mingzhu Group Co.	\\
600109	&	GJZQ	&	Sinolink Securities Co.	\\
600015	&	HXYH	&	Hua Xia Bank Co.	\\
600383	&	JDJT	&	Gemdale Corporation	\\
601328	&	JTYH	&	Bank Of Communications Co.	\\
600993 & MYL & Mayinglong Pharmaceutical Group Stock Co. \\
601009	&	NJYH	&	Bank Of Nanjing Co.	\\
600000	&	PFYH	&	Shanghai Pudong Development Bank Co.	\\
600604 & SBGX & Shanghai Shibei Hi-Tech Co. \\
600008	&	SCGF	&	Beijing Capital Co.	\\
600018	&	SGJT	&	Shanghai International Port (Group) Co.	\\
601088	&	SHE	&	China Shenhua Energy Company	\\
600009	&	SHJC	&	Shanghai International Airport Co.	\\
600841 & SHSC &Shanghai Diesel Engine Co. \\
600733 &   SQF & Chengdu Qianfeng Electronics Co. \\
600100	&	TFGF	&	Tsinghua Tongfang Co.	\\
600751 & TJHY & Tianjin Marine Shipping Co. \\
600058	&	WKFZ	&	Minmetals Development Co.	\\
600173	&	WLDC	&	Wolong Real Estate Group Co.	\\
601168	&	XBKY	&	Western Mining Co.	\\
600139	&	XBZY	&	Sichuan Western Resources Holding Co.	\\
600657	&	XDDC	&	Cinda Real Estate Co.	\\
600638	&	XHP	&	Shanghai New Huang Pu Real Estate Co.	\\
600755	&	XMGM	&	Xiamen International Trade Group Corp.	\\
600369	&	XNZQ	&	Southwest Securities Co.	\\
601166	&	XYYH	&	Industrial Bank Co.	\\
601766	&	ZGNC	&	CSR Corporation	\\
600890 & ZFGF & Cred Holding Co. \\
601628	&	ZGRS	&	China Life Insurance Company	\\
600028	&	ZGSH	&	China Petroleum\&Chemical Corporation	\\
601857	&	ZGSY	&	Petrochina Company	\\
601601	&	ZGTB	&	China Pacific Insurance (group) Co.	\\
601186	&	ZGTJ	&	China Railway Construction Corporation	\\
601390	&	ZGZT	&	China Railway Group	\\
600026	&	ZHFZ	&	China Shipping Development Company	\\
600036	&	ZSYH	&	China Merchants Bank Co.	\\
600030	&	ZXZQ	&	CITIC Securities Company	\\
\hline \hline
\end{tabular}
%\end{ruledtabular}
\end{table}

\end{appendix}

\end{document}